\documentclass[conference]{IEEEtran}
\usepackage[T1]{fontenc}
\usepackage{amsmath}
\usepackage{graphicx}
\usepackage[font=small,labelfont=bf]{caption}
\graphicspath{{images/}}
\usepackage[utf8]{inputenc}
\usepackage{algpseudocode}
\usepackage{algorithm}
\usepackage{multirow}               
\usepackage{color}
\usepackage{cite}
\usepackage[margin=1in]{geometry}
\usepackage{rotating}
\usepackage{framed}
\usepackage[table,xcdraw]{xcolor}
\def\quad{\hskip1em\relax}
\usepackage{amssymb}
\usepackage{url}
\usepackage{glossaries}

\title{Relation between Gene Content and Taxonomy in Chloroplasts}

\author{
\IEEEauthorblockN{Bashar Al-Nuaimi\IEEEauthorrefmark{1}\IEEEauthorrefmark{2}, Christophe Guyeux\IEEEauthorrefmark{1}, Bassam AlKindy\IEEEauthorrefmark{3}, Jean-Fran\c{c}ois Couchot\IEEEauthorrefmark{1}, and Michel~Salomon\IEEEauthorrefmark{1}}
\IEEEauthorblockA{\IEEEauthorrefmark{1}FEMTO-ST Institute, UMR 6174 CNRS, DISC Computer Science Department \\ Universit\'{e} de Bourgogne Franche-Comt\'{e}, France}
\IEEEauthorblockA{\IEEEauthorrefmark{2}Department of Computer Science, University of Diyala, Iraq}
\IEEEauthorblockA{\IEEEauthorrefmark{3}Department of Computer Science, University of Mustansiriyah, Iraq}

\IEEEauthorblockA{\IEEEauthorrefmark{0}christophe.guyeux@univ-fcomte.fr}
}

\begin{document}

\maketitle

\textbf{Abstract}
The aim of this study is to investigate the relation that can be found between the phylogeny of a large set of complete chloroplast genomes, and the evolution of gene content inside these sequences. Core and pan genomes have been computed on \textit{de novo} annotation of these 845 genomes, the former being used for producing well-supported phylogenetic tree while the latter provides information regarding the evolution of gene contents over time. It details too the specificity of some branches of the tree, when specificity is obtained on accessory genes. After having detailed the material and methods, we emphasize some remarkable relation between well-known events of the chloroplast history, like endosymbiosis, and the evolution of gene contents over the phylogenetic tree.


\section{Introduction}
Understanding the evolution of DNA molecules is a very complex problem, and no concrete and well established solution exists at present regarding the case of large DNA sequences. Our objective in this article is to start to show that this complex problem can be (at least partially) solved when considering genomes of reasonable size and who faced a rational number of recombination, like in the chloroplasts case. However, various difficulties remain to circumvent when dealing with such a specific case, and solving them require the design of new ad hoc tools. Candidates for such tools are presented in this article, and are applied on the chloroplast case.

Chloroplasts are one of the numerous types of organelles in the plant cell. The term of chloroplast comes from the combination of chloro and plastid, meaning that it is an organelle found in plant cells that contains the chlorophyll. Chloroplast has the ability to convert water, light energy, and carbon dioxide ($CO_2$) into chemical energy by using carbon-fixation cycle~\cite{uzman2003molecular}~(also called~\textit{Calven Cycle}, the whole process being called photosynthesis). This pivotal role explains why chloroplasts are at the basis of most trophic chains and are thus responsible for evolution and speciation.

Consequently, investigating the evolutionary history of chloroplasts is of great interest, and our long-term objective is to explore it by the mean of ancestral genomes reconstruction. This reconstruction will be achieved in order to discover how the molecules have evolved over time, at which rate, and to determine whether this way can present evidence of their cyanobacteria origin. This long-term objective necessitates numerous intermediate research advances. Among other things, it supposes to be able to apply the ancestral reconstruction on a well-supported phylogenetic tree of a representative collection of chloroplastic genomes. Indeed, sister relationship of two species must be clearly established before trying to reconstruct their ancestor. Additionally, it implies to be able to detect content evolution (modification of genomes like gene loss and gain) along this accurate tree. In other words, \textit{gene content evolution} on the one hand, and \textit{accurate phylogenetic inference} on the contrary, must be carefully regarded \textit{in the particular case of chloroplast sequences}, as the two most important prerequisites in our quest of the last universal common ancestor of these chloroplasts.

\begin{table}
\centering
\caption{Information on chloroplast sizes at highest taxonomic level}
\label{tab:genome size}
\begin{tabular}{|l|c|c|c|c|c|}
\hline
Taxonomy & nb. of & min length & max & average & standart \\
& genomes & length & length & & deviation \\
\hline
\hline
Alveolata & 4 & 85535 & 140426 & 115714.2 & 19648.3\\
Cryptophyta & 2 & 121524 & 135854 & 128689.0 & 7165.0\\
Euglenozoa & 7 & 80147 & 143171 & 98548.7 & 19784.5\\
Haptophyceae & 3 & 95281 & 107461 & 102683.6 &5307.6\\
Rhodophyta & 9 & 149987 & 217694 & 183755.5 &18092.2\\
Stramenopiles & 35 & 89599 & 165809 & 124895.1 & 15138.0\\
Viridiplantae & 775 & 80211 & 289394 & 150194.9 & 20376.8\\
\hline
\end{tabular}
\end{table}

The objective of this research work is to make significant progress in this quest, by providing material and methods required in the study of chloroplastic sequence evolution. Contributions of this article consist in the computation of core and pan genomes of the 845 complete genomes available on the NCBI, in the production of a well-supported phylogenetic tree based on core sequences as large as possible, and on the study of the produced data. In particular, we start to emphasize some links between the phylogenetic tree and evolution of gene content.

The paper is structured as follows. In the next section, material and methods applied in this study are presented, which encompass genome acquisition and annotation, core and pan genome analysis, and phylogenetic investigations. Obtained results related to such analyzes are detailed in Section~\ref{sec:obtainedRes}, on the chloroplast case.
This article ends with a conclusion section, in which the study is summarized and intended future work is outlined.

\section{Materials and methods}
\label{Data acquisition}

\subsection{Data acquisition}
A set of 845 chloroplastic genomes (green algae, red algae, gymnosperms, and so on) has been downloaded from the NCBI website, representing all the available complete genomes at the date of March, 2016 (see Table~\ref{tab:genome size}). An example of such sequences, taken from the \textit{Streptophyta} clade (a \textit{Viridiplantae}), is provided in Table~\ref{tab:Streptophyta}. Note that this set does not really constitute a very balanced representation of the diversity of plants, as plants of particular and immediate interest to us like \textit{Viridiplantae} are first sequenced. We must however deal with such bias, as genomic data acquisition is most of the time human-centred. This set of sequences presents too a certain variability in terms of length, as detailed in Table~\ref{tab:genome size}.

\begin{table}
\caption{Example of genomes information of~\textit{Streptophyta}~clade}
\begin{center}
\begin{tabular}{l|c|c|c}
\hline
Organism name& Accession &  Sequence & Nb of  \\
 & number &  length & CDS\\
\hline\hline
\textit{Epimedium sagittatum} & NC\_029428.1  & 158273 & 85  \\ 
\textit{Berberis bealei} & NC\_022457.1  & 164792 & 267  \\ 
\textit{Torreya fargesii} & NC\_029398.1 & 137075 & 100  \\ 
\textit{Lepidozamia peroffskyana} & NC\_027513.1 &  165939 & 93  \\ 
\textit{Actinidia chinensis} & NC\_026690.1 &  156346 & 271  \\ 
\textit{Quercus aliena} & NC\_026790.1  & 160921  & 259  \\ 
\textit{Quercus aquifolioides} & NC\_026913.1 & 160415 & 176  \\ 
\textit{Sedum sarmentosum} & NC\_023085.1 & 150448 & 99  \\  \hline\hline
\end{tabular}
\end{center}
\label{tab:Streptophyta}
\end{table}

Each genome has been annotated with DOGMA~\cite{RDogma}, an online automatic and accurate annotation tool of organellar genomes, following a same approach than in~\cite{Alkindy2014}. To apply it on our large scale, we have written (with the agreement of DOGMA authors) a script that automatic send requests to the website. By doing such annotations, the same gene prediction and naming process has been applied with the same average quality of annotation. In particular, when a gene appears twice in the considered set of genomes, it receives twice the same name (no spelling error). At this level, each genome is then described by an ordered list of gene names, with possible duplications (other approaches for the annotation stage are possible, see, \textit{e.g.},~\cite{Alkindy2014}). This description will allow us to investigate, later in this article, the evolution of gene content among the species tree, leading to the study of core and pan genomes recalled below.

\subsection{Core and pan genome}
\label{Core and pan genome}
Given a collection of genomes, it is possible to define their core genes as the common genes that are shared among all the species, while the pan genome is the union of all the genes that are in at least one genome (\emph{all} the species have each core gene, while a pan gene is in \emph{at least one} genome).
Shared genes are evidences of evolution from a common ancestor and of the relatedness of chloroplast organisms. 

\begin{table}
\caption{Summarized properties of the pan genomes at the highest taxonomic level.}
\label{taxo1}
\centering
\begin{tabular}{|c|c|c|c|c|}
\hline
\textit{Taxonomy}&Nb.&Min N.b&Max N.b&Average Nb. \\
 &genomes&of pan genes& of pan genes&of pan genes\\ 
\hline
\hline
\textit{Alveolata} &  4 & 253 & 266 & 262.25 \\
\hline
\textit{Cryptophyta} &  2 & 258 & 259 & 258.5 \\
\hline
\textit{Euglenozoa} &  7 & 193 & 267 & 253.428 \\
\hline
\textit{Haptophyceae} &  3 & 251 & 266 & 258.333 \\
\hline
\textit{Rhodophyta} &  9 & 156 & 267 & 246.222\\
\hline
\textit{Stramenopiles} &  35 & 73 & 271 & 238.971 \\
\hline
\textit{Viridiplantae} &  775 & 85 & 271 & 229.827 \\
\hline
\end{tabular}
\end{table}

To distinguish and determine the core genes may be of importance either to identify the specificity and the shared functionality of a given set of species, or to evaluate their phylogeny using the largest set of shared coding sequences. In the case of chloroplasts, an important category of genome modification is indeed the loss of functional genes, either because they become ineffective or due to transfer to the nucleus. Thereby, a small number of gene loss among species may indicate that these species are close to each other and belong to a similar lineage, while a significant loss means distant lineages.

So core genome is obviously of importance when inferring the phylogenetic relationship, while accessory genes of pan genome explain in some extend each species specificity. We have formerly proposed three approaches for eliciting core genomes. The first one uses correlations computed on predicted coding sequences~\cite{Alkindy2014}, while the second one uses all the information provided during an accurate annotation stage~\cite{Alkindy_BIBM2014}. The third method takes the advantages from the first two approaches, by considering gene information and DNA sequences, in order to find the targeted core genome~\cite{Alkindy_IWWBIO}.

\begin{figure}[H]
\begin{center}
    \includegraphics[width=0.50\textwidth]{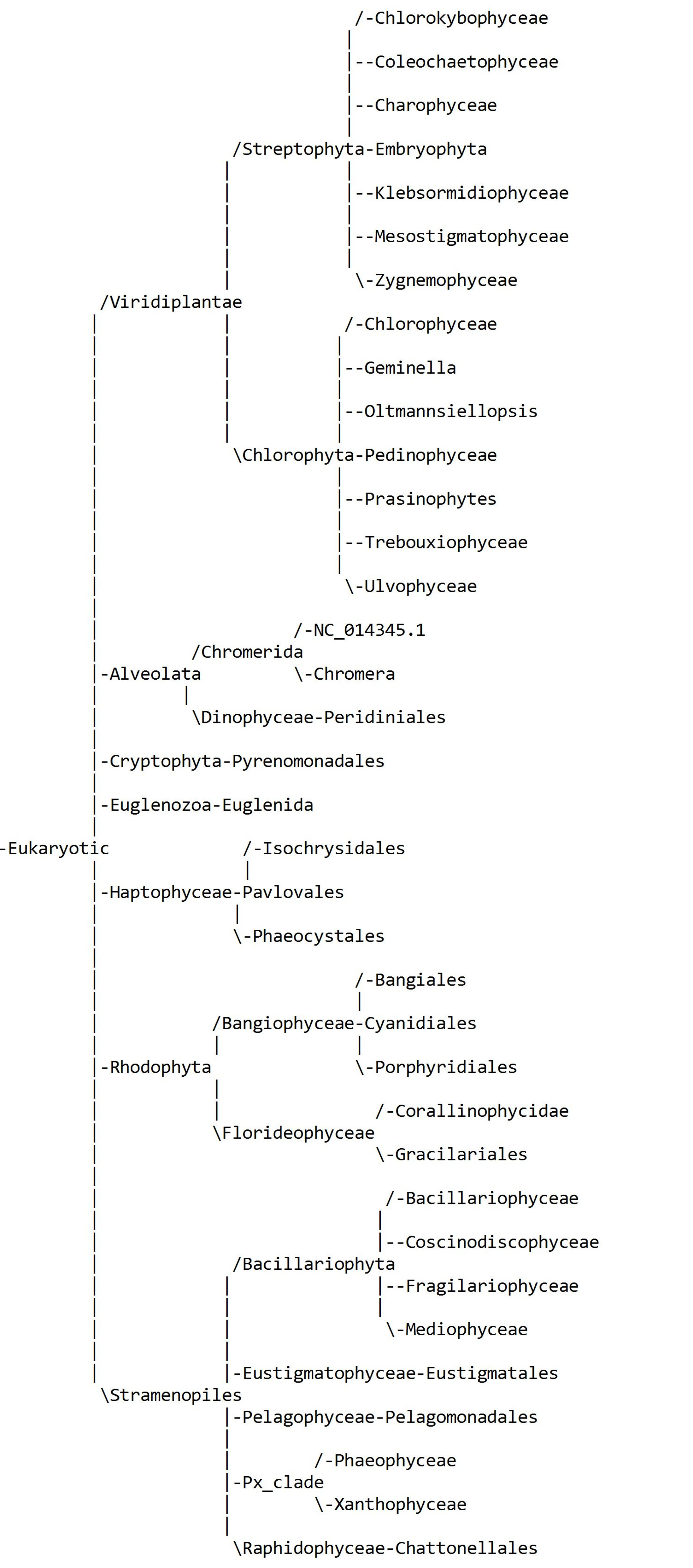}
    \caption{Phylogenetic tree overview\label{fig:final}}
\end{center}
\end{figure}

We have found the core genome\footnote{\begin{color}{red}All data are available at...\end{color}} of each selected family by using the second method described in the previous paragraph~\cite{Alkindy_BIBM2014}. Obtained results regarding gene content are discussed in the next section. The core genome has been used too for our phylogenetic investigation of chloroplast sequences, which has been applied as described hereafter.

\subsection{Phylogeny study}
The next step when trying to reconstruct the evolution of gene content over time is to deeply investigate the phylogeny of these chloroplasts, in order to obtain a tree as supported as possible. Indeed, a branching error in the tree may lead to an erroneous transmission of an ancestral state, which is dramatically perpetuated until reaching the last universal common ancestor. However, as we considered all existing plant taxa, we faced chloroplastic sequences that have diverged a lot since two billion of years, so the core genome of these 845 sequences is very small when compared with sequence length of each representative, and inferring a tree on such a partial information will probably lead to numerous errors.

The approach that has been regarded in our study was then to group the plant families per close packets (same family in the taxonomy). Such grouping has enlarged the number of shared gene sequences (core genes of the considered family) on which a more representative phylogeny can be computed~\cite{DBLP:journals/corr/AlKindyGCSPB15}. 
After having aligned the core genes of each family using MUSCLE~\cite{edgar2004muscle} on our supercomputer facilities, we then have inferred a phylogenetic tree per family.

\begin{figure}
\begin{center}
    \includegraphics[width=0.5\textwidth]{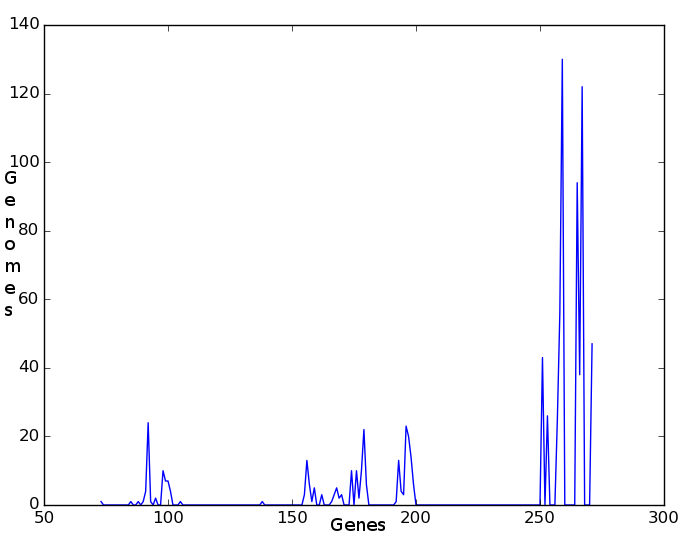}
    \caption{The distributions of chloroplast genomes depending on the genomes size.\label{fig:genes}}
\end{center}
\end{figure}

To obtain such a tree, the RAxML~\cite{stamatakis2014raxml,rizzo2007review} program has been employed to compute the phylogenetic maximum-likelihood (ML) function with the setup described hereafter. General Time Reversible model of nucleotide substitution, with $\Gamma$ model of rate heterogeneity and hill-climbing optimization method. The \textit{Prochlorococcus marinus} (NC\_009091.1) cyanobacteria species has finally been chosen as outgroup, due to the supposed cyanobacteria origin of chloroplasts.

After such a computing, if all bootstrap values are larger than 95\%, then we have consider that the phylogeny is resolved, as the largest possible number of genes has led to a very well supported tree. In case where some branches are not supported, we can wonder whether a few genes can be incriminated in this lack of assistance, for a large variety of reasons encompassing homoplasy, stochastic errors, undetected paralogy, incomplete lineage sorting, horizontal gene transfers, or even hybridization. Such problem has been resolved by finding the largest subset of core genes leading to the most supported tree, by the heuristic approach coupled with statistical LASSO tests described in~\cite{alsrraj2015well,DBLP:journals/corr/AlKindyGCSPB15}.
Obtained trees are then merged on a well-supported and representative supertree.

\section{Obtained results}
\label{sec:obtainedRes}

\subsection{Phylogenetic investigations}
The approach detailed in the previous section has led to a well supported phylogenetic tree of the whole available chloroplasts, with the ordered list of genes at each leaf of the tree. An overview of the latter is provided in Figure~\ref{fig:final}. Obtained tree available too on our website is in general coherent with the NCBI taxonomy, except in some specific locations.

By going into the details of the obtained tree, it is well known that the first plants endosymbiosis ended in a great diversification of lineages comprising Red Algae, Green Algae, and Land Plants (terrestrial). The interesting point in the production of our results is that the organisms resulting from the first endosymbiosis are distributed in each of the lineages found in the chloroplast genome structure evolution as outlined in Figure~\ref{fig:final}.

More precisely, all Red Algae chloroplasts are grouped together in one lineage, while Green Algae and Land Plant chloroplasts are all in a second lineage. 
Furthermore, organisms resulting from the secondary endosymbioses, as listed in Table~\ref{taxo2}, are well localized in the tree: both the chloroplasts of Brown Algae and~\textit{Dinoflagellates} representatives are found exclusively in the lineage also comprising the Red Algae chloroplasts from which they evolved, while the~\textit{Euglens} is related to Green Algae from which they evolved. This latter makes sense regarding biology, history of lineages, and theories of chloroplasts origins (and so photosynthetic ability) in different~\textit{Eucaryotic} lineages~\cite{li2013complete}.

\subsection{Gene content}
Let us now investigate the gene content level of the tree. Indeed, genes are rearranged in the genome by evolutionary events like insertion, deletion, transposition, and inversion, which are called genome rearrangements~\cite{suyama2001evolution}. Such rearrangements can be studied, considering that we have both the gene contents and the phylogeny. A general overview of obtained results, in terms of gene contents (pan genome) evolution at the top taxonomy level, is provided in Table~\ref{taxo1}, and it is detailed for the following taxonomic level in Table~\ref{taxo2}.

The core genome is constituted by 36 coding sequences, namely: \textit{ATPA, ATPB, ATPH, ATPI, PETB, PETG, PSAA, PSAB, PSAC, PSAJ, PSBA, PSBC, PSBD, PSBE, PSBF, PSBH, PSBI, PSBJ, PSBL, PSBN, PSBT, PSI\_PSBT, RBCL, RPL14, RPL16, RPL2, RPL20, RPL36, RPS11, RPS12, RPS12\_3END, RPS14, RPS19, RPS2, RPS7}, and \textit{RRN16}. The pan genome of the whole considered species, for its part, contains 268 genes. Note that, according to our computation, no gene was specific to a given clade (that is, present in only one clade).

\begin{table*}
\caption{Taxonomy in the second level}
\label{taxo2}
\centering
\begin{tabular}{|l|l|c|c|c|c|}
\hline
\textit{Taxonomy}& &Nb.&Min N.b&Max N.b& Avg N.b\\
& &genomes&of pan genes&of pan genes& of pan genes\\
\hline
\hline
\textit{Alveolata}&\textit{Chromerida}&2&253&265& 259.0 \\
&\textit{Dinophyceae}& 2 & 265 & 266 & 265.5 \\
\hline
\textit{Cryptophyta}&\textit{Pyrenomonadales}& 2 & 258 & 259 & 258.5 \\
\hline
\textit{Euglenozoa}&\textit{Euglenida}& 7 & 193 & 267 & 253.428 \\
\hline
\textit{Haptophyceae}&\textit{Phaeocystales}& 1 & 266 & 266 & 266.0 \\
&\textit{Isochrysidales}  & 1 & 251 & 251 & 251.0 \\
&\textit{Pavlovales}& 1 & 258 & 258 & 258.0 \\
\hline
\textit{Rhodophyta}&\textit{Bangiophyceae}& 6 & 156 & 266 & 240.166 \\
&\textit{Florideophyceae}& 3 & 251 & 267 & 258.333 \\
\hline
\textit{Stramenopiles}&\textit{Px\_clade}& 6 & 251 & 271 & 261.166 \\
&\textit{Bacillariophyta}& 20 & 138 & 271 & 231.35 \\
&\textit{Eustigmatophyceae}& 6 & 253 & 267 & 262.16 \\
&\textit{Raphidophyceae}& 1 & 258 & 258 & 258.0 \\
&\textit{Pelagophyceae}& 2 & 73 & 266 & 169.5 \\
\hline
\textit{Viridiplantae}&\textit{Chlorophyta}& 58 & 156 & 271 & 244.517 \\
 &\textit{Streptophyta}& 717 & 85 & 271 & 228.638 \\
\hline
\end{tabular}
\end{table*}

\subsection{Relations between gene content and phylogeny}
\begin{figure}
\begin{center}
    \includegraphics[width=0.5\textwidth]{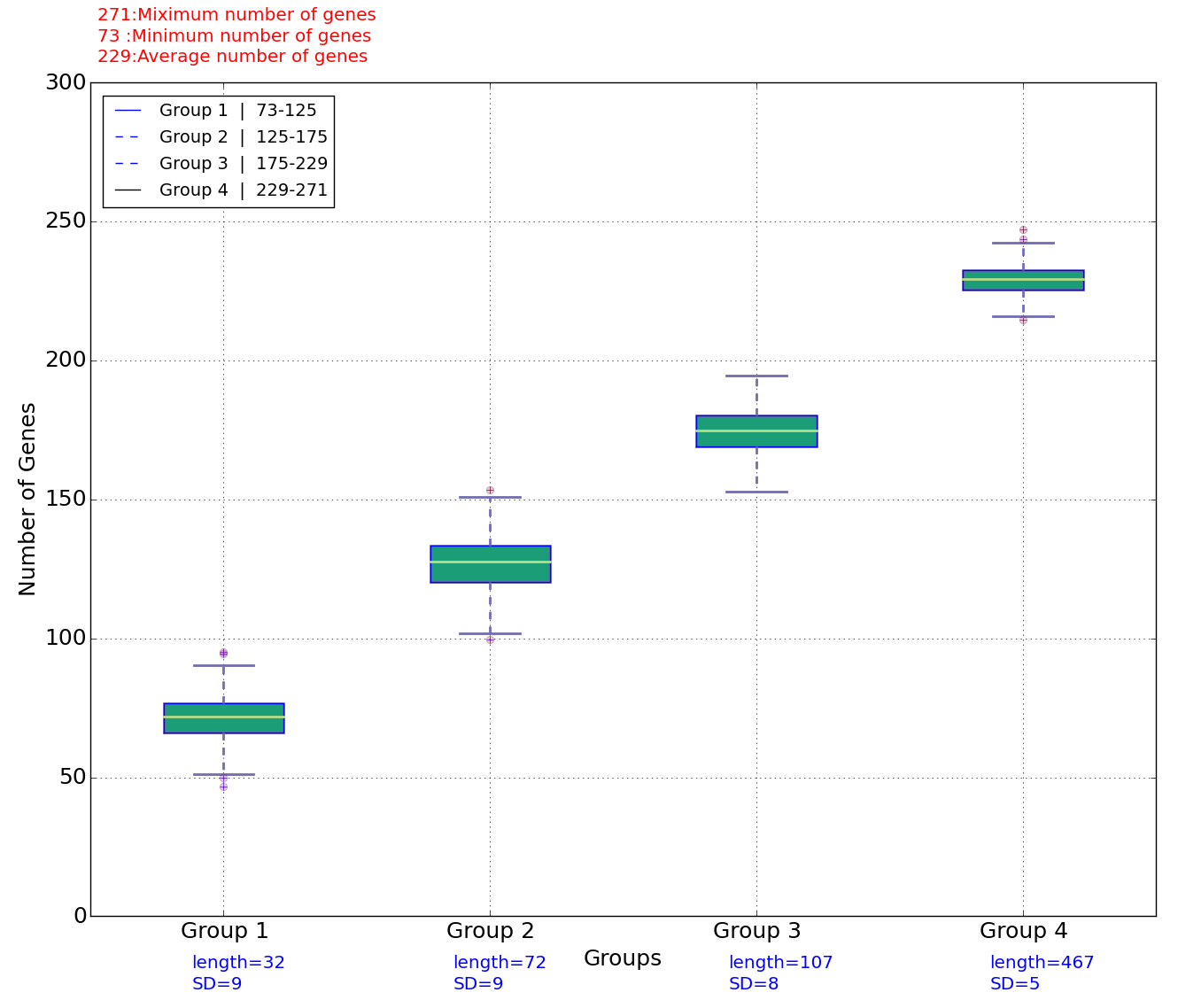}
    \caption{Classification of chloroplast genomes according to numbers of pan genes.\label{fig:pan_genomes}}
\end{center}
\end{figure}

We then have further investigated the distribution of number of genes according to the group of species. Obtained results are reproduced in Figures~\ref{fig:genes} and~\ref{fig:pan_genomes}. Four groups have appeared among the 845 genomes, which are taxonomically coherent. 
As shown in Fig.~\ref{fig:pan_genomes}, the cluster of largest genomes has a number of genes ranging from 229 to 271, while in the group of smallest genomes, the lowest number of genes is for the \textit{Viridiplantae} case. 
In particular, among the genomes having less than 120 genes, we found accession number NC\_012903.1 (\textit{Eukaryota, Stramenopiles, Pelagophyceae, Pelagomonadales, Aureoumbra
lagunensis}), and 63 \textit{Spermatophyta} species: 3 \textit{Pinidae}, 58 \textit{Magnoliophyta}, one \textit{Cycadidae}, and finally one \textit{Gnetidae}.
We finally obtain chloroplast genomes varying from 73 to 271 genes.

We can further note that (1) most of the organisms in green lineage (green algae and land plants) have a lower number of genes in their chloroplasts compared to the red algae. (2) Most land plants have genome sizes ranging between 120 and 160 kb~\cite{green2011chloroplast}. (3) Most of the differences in genome size are due to the number of paralogous genes.

\begin{figure*}
    \centering
    \includegraphics[scale=0.115]{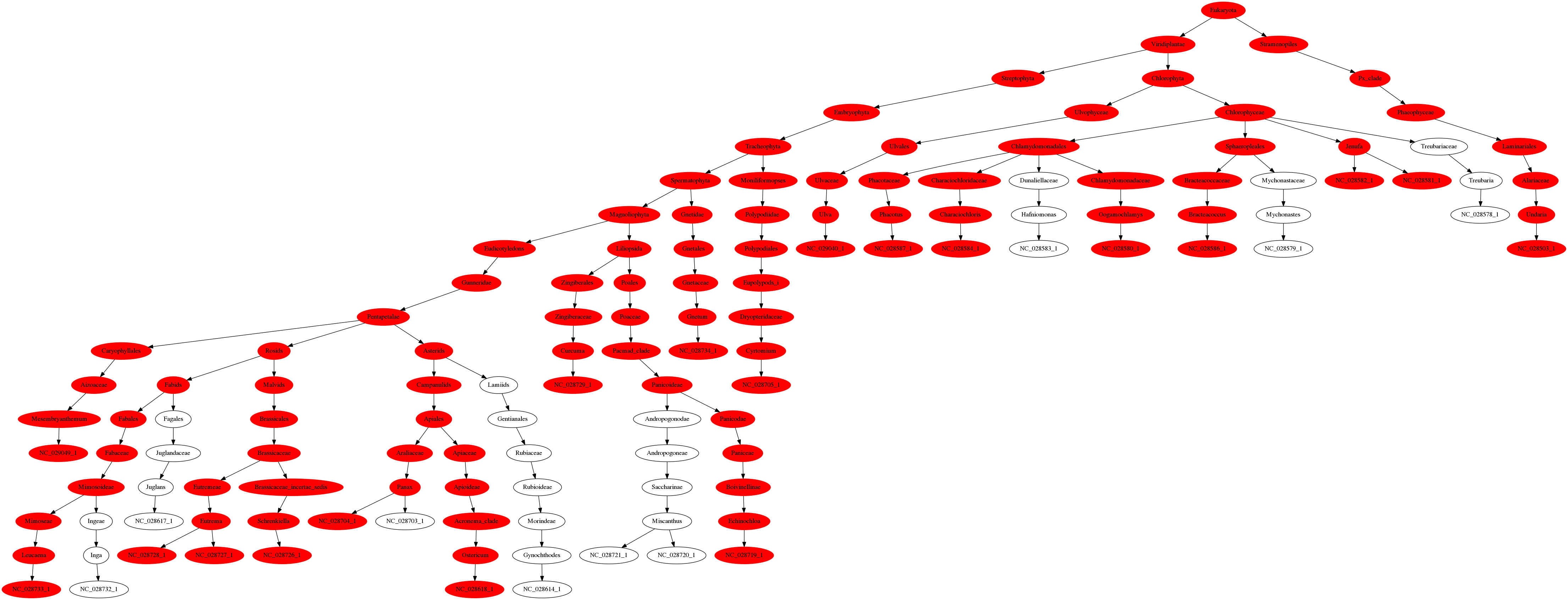}
    \caption{ACCA gene loss in various branches of the tree}
    \label{fig:ACCApan}
\end{figure*}

When regarding more deeply the ordered list of genes to investigate the reasons of such differences of size, it appears to us that the gene content evolution can mostly be explained by repetitions of some genes and the loss of other ones: no large scale recombination is responsible of such variations. Usual case is as in Figure~\ref{fig:ACCApan} for ACCA pan gene, on which single vulnerable genes are lost, possibly in various independent branches, due to deletere mutations. Such results have been obtained by comparing, for each couple of close genomes, all gene names and positions, by practicing a naked eye investigation using homemade scripts. Some mutation and indel events are provided too in Table~\ref{my-label}, for the sake of illustration.

\begin{table*}
\resizebox{0.9\textwidth}{!}{\begin{minipage}{\textwidth}
\centering
\caption{Example of comparison between pairwise genomes from various species, to investigate the changes that occurred within branches of the tree.}

\label{my-label}
\begin{tabular}{|c|c|c|c|c|c|c|c|}
\hline
\rowcolor[HTML]{C0C0C0} 
\multicolumn{1}{|r|}{\cellcolor[HTML]{C0C0C0}{\color[HTML]{000000} Index}} & {\color[HTML]{000000} Clade}             & {\color[HTML]{000000} Sub-kingdom}    & {\color[HTML]{000000} \begin{tabular}[c]{@{}c@{}}Order/ Family\end{tabular}} & {\color[HTML]{000000} \begin{tabular}[c]{@{}c@{}}Genome \\ name\end{tabular}} & {\color[HTML]{000000} \begin{tabular}[c]{@{}c@{}}N.b of \\ pan genes\end{tabular}} & {\color[HTML]{000000} \begin{tabular}[c]{@{}c@{}}Deletion/\\ Insertion\end{tabular}} & {\color[HTML]{000000} \begin{tabular}[c]{@{}c@{}}Matching \\ ratio\end{tabular}} \\ \hline

 &   &    &     & \textit{Camelina}  & 92   &   &    \\ \cline{5-6}
\multirow{-2}{*}{1}  & \multirow{-2}{*}{\textit{Viridiplantae}} & \multirow{-2}{*}{\textit{Embryophyta}}   &\multirow{-2}{*}{\textit{Camelineae}}  & \textit{Barbarea} & 267  & \multirow{-2}{*}{173/0} & \multirow{-2}{*}{48.46} \\ \hline
&   &   &    & \textit{Aquilaria}   & 101   &  & \\ \cline{5-6} \multirow{-2}{*}{2} & \multirow{-2}{*}{\textit{Viridiplantae}} & \multirow{-2}{*}{\textit{Embryophyta}}  & \multirow{-2}{*}{\textit{Camelineae}} & \textit{Hibiscus}  & 267                               & \multirow{-2}{*}{164/0} & \multirow{-2}{*}{28.26}  \\ \hline &  &   &  & \textit{Acer}  & 92   &   &   \\ \cline{5-6}
\multirow{-2}{*}{3}  & \multirow{-2}{*}{\textit{Viridiplantae}} & \multirow{-2}{*}{\textit{Embryophyta}}  &\multirow{-2}{*}{\textit{Sapindales}}& \textit{Azadirachta} & 267  & \multirow{-2}{*}{173/0}  & \multirow{-2}{*}{49.02}   \\ \hline
 &   &    & & \textit{Lepidozamia}   & 92    &  &  \\ \cline{5-6} \multirow{-2}{*}{4} & \multirow{-2}{*}{\textit{Viridiplantae}} &\multirow{-2}{*}{\textit{Embryophyta}}   & \multirow{-2}{*}{\textit{Zamiaceae}} & \textit{Zamia}  & 267  & \multirow{-2}{*}{172/0}  & \multirow{-2}{*}{51.66}  \\ \hline &      &   &  & \textit{Lavanduleae}    & 92    &   &   \\ \cline{5-6}
\multirow{-2}{*}{5} & \multirow{-2}{*}{\textit{Viridiplantae}} & \multirow{-2}{*}{\textit{Embryophyta}}   & \multirow{-2}{*}{\textit{Lamiaceae}} & \textit{Perman} & 267  & \multirow{-2}{*}{173/0} & \multirow{-2}{*}{49.91}  \\ \hline
 &       &        &  & \textit{Aureoumbra}   & 73  & &  \\ \cline{5-6}
\multirow{-2}{*}{6}  & \multirow{-2}{*}{\textit{Stramenopiles}} & \multirow{-2}{*}{\textit{Pelagophyceae}} & \multirow{-2}{*}{\textit{Pelagomonadales}}  & \textit{Aureococcus}  & 267  & \multirow{-2}{*}{193/0}  & \multirow{-2}{*}{27.13}     \\ \hline
&       &        &  & \textit{Epimedium}   & 85  & &  \\ \cline{5-6}
\multirow{-2}{*}{7}  & \multirow{-2}{*}{\textit{Viridiplantae}} & \multirow{-2}{*}{\textit{Eudicotyledons}} & \multirow{-2}{*}{\textit{Berberidoideae}}  & \textit{berberis}  & 267  & \multirow{-2}{*}{180/0}  & \multirow{-2}{*}{33.52}     \\ \hline
&       &        &  & \textit{NC\_026690\_1}   & 92  & &  \\ \cline{5-6}
\multirow{-2}{*}{8}  & \multirow{-2}{*}{\textit{Viridiplantae}} & \multirow{-2}{*}{\textit{Eudicotyledons}} & \multirow{-2}{*}{\textit{Actinidia}}  & \textit{NC\_026691\_1}  & 271  & \multirow{-2}{*}{174/0}  & \multirow{-2}{*}{48.63}     \\ \hline

\end{tabular}
\end{minipage}}
\end{table*}

\section{Conclusion}\label{Conclusion}
In this article, we made significant progress in the study of chloroplastic sequence evolution, by providing material and methods required in the quest of the ancestral genome of the chloroplasts. A large set of complete chloroplast genomes has been studied \textit{de novo} regarding both core and pan genomes, phylogenetic relationship, and gene content modifications. We then started to study the produced data, by emphasizing some remarkable relations between well-known events of the chloroplast history and the evolution of gene contents over the phylogenetic tree.

In future work, our intention is to investigate more systematically such relations between remarkable ancestral nodes in the tree, endosymbiosis events, and evolution of gene content. We will wonder whether some branches of the trees are statistically remarkable when considering gene content (for instance, do we have a correlation between the presence or absence of a subset of genes, and a particular taxonomy). Then, the gene ordering and content of each ancestral node will be computed using ad hoc algorithms, ancestral DNA sequences will be inferred, and ancestral intergenic regions will be deduced, in order to have all ancestral genomes with confidence indications like probabilities. The produced ancestral genomes will then be used to investigate hypotheses formulated by biologists, regarding the origin of chloroplasts, their recombination events, and the transfer of some material to the nucleus. We will in particular study whether recombination events where uniform over time and on the whole sequence, or if it is possible to highlight some hot spots of recombination in the history of these chloroplasts.

\medskip

\textit{All computations have been performed using the ``Mésocentre de calcul de l'Université de Franche-Comté'' supercomputer facilities.}

\bibliographystyle{unsrt}
\bibliography{Ref}

\end{document}